\documentstyle[aps,prl,twocolumn,epsfig]{revtex}
\begin{document}
\newcommand{\TGLOB}{T_{\rm glob}}
\newcommand{\TLOC}{T_{\rm loc}}
\newcommand{\EGLOB}{E_{\rm glob}}
\newcommand{\ELOC}{E_{\rm loc}}

\twocolumn[\hsize\textwidth\columnwidth\hsize\csname @twocolumnfalse\endcsname

\title{\bf Efficient Behavior of Small-World Networks}
\author{ Vito Latora$^{1,2}$ and Massimo Marchiori$^{3,4}$
\\
$^1$ Laboratoire de Physique Th\'eorique et Mod\`eles Statistiques,
        Universit\'e Paris-Sud, \\
    Bat.100, 91405 Orsay Cedex, France\\
           $^2$ Department of Physics and Astronomy,
University of Catania, and INFN, Italy\\
           $^3$ W3C and Lab. for Computer Science,
Massachusetts Institute of Technology, USA\\
           $^4$ Department of Computer Science, University of Venice, Italy }
\date{\today}
\maketitle

\begin{abstract}
We introduce the concept of {\em efficiency\/} of a network,
measuring how efficiently it exchanges information. By using this
simple measure small-world networks are seen as systems that are
both globally and locally efficient. 
This allows to give a clear physical meaning to the concept 
of small-world, and also to
perform a precise quantitative analysis of both weighted and unweighted 
networks. We study
neural networks and man-made communication and transportation
systems and we show that the underlying general principle of their
construction is in fact a small-world principle of high
efficiency.

PACS numbers 89.70.+c, 05.90.+m, 87.18.Sn, 89.40.+k
\end{abstract}

\vspace{0.5cm}

   ] 

We live in a world of networks. In fact any complex system in nature
can be modeled as a network, where vertices are the elements of the system
and edges represent the interactions between them. Coupled biological and
chemical systems, neural networks,
social interacting species, computer networks or the Internet
are only few of such examples\cite{yaneer}.
Characterizing the structural properties of the networks is then
of fundamental importance to understand the complex dynamics of
these systems. A recent paper \cite{watts} has shown that
the connection topology of some biological and social networks
is neither completely regular nor completely random.
These networks, there named {\it small-worlds}, in analogy with the concept of
small-world phenomenon developed 30 years ago in social psychology
\cite{milgram}, are in fact highly clustered like regular lattices,
yet having small characteristics path lengths like random graphs.
The original paper has triggered a large interest in the study
of the properties of small-worlds (see ref. \cite{newman1} for a
recent review).
Researchers have focused their attention on different aspects:
study of the inset mechanism \cite{barrat,marchiori,amaral1},
dynamics \cite{lago} and spreading of diseases on small-worlds \cite{newman2},
applications to social networks \cite{newman3,amaral2} and to
the Internet \cite{barabasi1,barabasi2}.
In this letter we introduce the concept of {\em efficiency\/} of a
network, measuring how efficiently information is exchanged over the
network. By using efficiency small-world networks results as systems
that are both globally and locally efficient.
This formalization gives a clear physical meaning to the concept
of small-world, and also allows a precise quantitative
analysis of unweighted and weighted networks.
We study several systems, like brains,
communication and transportation networks,
and show that the underlying general principle of their
construction is in fact a small-world principle,
provided attention is taken not to ignore an important observational
property (closure).

\noindent
We start by reexamining the original
formulation proposed in ref. \cite{watts}.
There, a generic graph $\bf G$ with $N$ vertices and $K$ edges is considered.
$\bf G$ is assumed to be
{\it unweighted}, i.e.\ edges are all equal,
{\it sparse} ($K \ll N(N-1)/2$), and
{\em connected\/}. i.e.\ there exists at
least one path connecting any couple of vertices with a finite number
of steps.
$\bf G$ is therefore represented by simply giving the adjacency
(or connection) matrix, i.e.\
the $N \cdot N$ matrix whose entry $a_{ij}$ is $1$ if there is an edge joining
vertex $i$ to vertex $j$, and $0$ otherwise.
An important quantity of $\bf G$ is the degree of vertex $i$, i.e.\ the
number $k_i$ of edges incident with vertex $i$ (the number
of neighbours of $i$). The average value of
$k_i$ is $k=2K/N$.
Once $\{a_{ij}\}$ is given it can be used to calculate
the matrix of the shortest path lengths $d_{ij}$ between two generic
vertices $i$ and $j$.
The fact that $\bf G$ is assumed to be connected implies that
$d_{ij}$ is positive and finite $\forall i \neq j$.
In order to quantify the structural properties of $\bf G$,
\cite{watts} proposes to evaluate two different
quantities: the characteristic path length $L$ and
the clustering coefficient $C$.
$L$ is the average distance between two generic vertices
$L= \frac {1}{N(N-1)} \sum_{i\neq j} d_{ij}$, and $C$
is a local property defined as
$C= \frac {1}{N} \sum_{i} C_i$. Here $C_i$ is the number of
edges existing in $\bf G_i$, the subgraph
of the neighbors of i, divided by the maximum possible
number $k_i(k_i-1)/2$.
In \cite{watts} a simple method is considered to produce a
class of graphs with increasing randomness.
The initial graph  $\bf G$ is taken to be a one-dimensional lattice
with each vertex connected to its k neighbours and with periodic
boundary conditions. Rewiring each edge at random with probability $p$,
$\bf G$ can be tuned in a continuous way from a regular lattice ($p=0$)
into a random graph ($p=1$).
For the regular lattice we expect $L\sim N/2k$ and a high
clustering coefficient $C= 3/4 (k-2)/(k-1)$,
while for a random graph $L \sim \ln N /ln(k-1)$ and $C\sim k/N$
\cite{bollobas,barrat}.
Although in the two limit cases a large $C$ is associated
to a large $L$ and vice versa a small $C$ to a small $L$,
the numerical experiment reveals an intermediate regime at
small $p$ where the system is highly clustered like regular lattices,
yet having small characteristics path lengths like random graphs.
This behavior is there called small-world and it is found to be
a property of some social and biological networks analyzed
\cite{watts}.

Now we propose a more general set-up to investigate real
networks. We will show that:

\noindent
{\bf -}
the definition of small-world behavior can be given in terms of a
single variable with a physical meaning, the {\it efficiency} $E$
of the network.

\noindent
{\bf -}
$1/L$ and $C$ can be seen as first approximations of $E$ evaluated
resp.\ on a global and on a local scale.

\noindent
{\bf -} we can drop all the restrictions on the system, like
unweightedness, connectedness and sparseness.

\noindent
We represent a real network as a generic {\it weighted} (and possibly
even {\em non sparse\/} and {\it non connected}) graph  $\bf G$.
Such a graph needs two matrices to be described:
the adjacency matrix $\{a_{ij}\}$ defined as for the unweighted graph,
and the matrix $\{\ell_{ij}\}$ of physical distances.
The number $\ell_{ij}$ can be the space distance between
the two vertices or the strength of their possible
interaction: we suppose $\ell_{ij}$ to be known even if in
the graph there is no edge between $i$ and $j$.
To make some examples, $\ell_{ij}$ can be the geographical distance
between stations in transportation systems
(in such a case $\ell_{ij}$ respects the triangle equality,
though this is not a necessary assumption), the time
taken to exchange a packet of information between routers in the
Internet, or the inverse velocity of chemical reactions along
a direct connection in a biological system.
Of course, in the particular case of an unweighted graph
$\ell_{ij}=1 ~\forall i \neq j$.
\noindent
The shortest path length $d_{ij}$ between two generic points
$i$ and $j$ is the smallest sum of the physical distances
throughout all the possible paths in the graph from $i$ to $j$.
The matrix $\{d_{ij}\}$ is therefore calculated by using
the information contained both in matrix $\{a_{ij}\}$
and in matrix $\{\ell_{ij}\}$.
We have $d_{ij} \ge \ell_{ij} ~\forall i,j$, the equality
being valid when there is an edge between $i$ and $j$.
Let us now suppose that the system is parallel, i.e.\
every vertex sends information concurrently
along the network, through its edges.
The efficiency $\epsilon_{ij}$ in the communication between
vertex $i$ and $j$ can be then defined to be inversely proportional
to the shortest distance:  $\epsilon_{ij} = 1/d_{ij} ~\forall i,j$.
When there is no path in the graph
between $i$ and $j$, $d_{ij}=+\infty$ and consistently
$\epsilon_{ij}=0$.
The average {\it efficiency} of $\bf G$ can be defined as:
\begin{equation}
\label{efficiency}
E({\bf G})=
\frac{ {{\sum_{{i \ne j\in {\bf G}}}} \epsilon_{ij}}  } {N(N-1)}
          = \frac{1}{N(N-1)}
{\sum_{{i \ne j\in {\bf G}}} \frac{1}{d_{ij}}}
\end{equation}
To normalize $E$ we consider the ideal case
${\bf {G_{id}}}$ in which the graph $\bf G$ has all the
$N(N-1)/2$ possible edges.
In such a case the information is propagated in the most efficient
way since $d_{ij} =\ell_{ij}~\forall i,j$, and $E$
assumes its maximum value $E({\bf {G_{id}}})= \frac{1}{N(N-1)}
{\sum_{{i \ne j\in {\bf G}}} \frac{1}{l_{ij}}}$.
The efficiency $E({\bf G})$ considered in the following of the paper
is always divided by $E({\bf {G_{id}}})$ and
therefore $0 \le E({\bf G}) \le 1$.
Though the equality $E=1$ is valid when there is an edge between each
couple of vertices, real networks can reach a high value of $E$.

In our formalism, we can define the small-world behaviour
by using the single measure $E$ to analyze both the local and global behaviour,
rather than two different variables $L$ and $C$.
The quantity  in eq. (\ref{efficiency}) is the {\it global efficiency}
of $\bf G$ and we therefore name it $\EGLOB$.
Since $E$ is defined also for a disconnected graph we can
characterize the local properties of $\bf G$
by evaluating for each vertex $i$ the efficiency of
$\bf {G_i}$, the subgraph of the neighbors of $i$.
We define the {\it local efficiency} as the average efficiency
of the local subgraphs,
$\ELOC = 1/N  \sum_{i \in {\bf G}} ~  E(\bf {G_i})$. This quantity
plays a role similar to the clustering coefficient $C$.
Since $i \notin \bf {G_i}$, the local efficiency $\ELOC$ tells how much
the system is {\it fault tolerant}, thus how efficient
is the communication between the first neighbours of $i$ when $i$
is removed \cite{faultbarabasi}.
The definition of small-world can now be rephrased and generalized
in terms of the information flow:
small-world networks have high $\EGLOB$ and $\ELOC$,
i.e.\ are very efficient in global and local communication.
This definition is valid both for unweighted and weighted
graphs, and can also be applied to disconnected and/or non sparse graphs.

It is interesting to see the correspondence between our measure
and the quantities $L$ and $C$ of \cite{watts} (or,
correspondingly, $1/L$ and $C$). The fundamental difference is
that $\EGLOB$ is the efficiency of a {\em parallel systems\/},
where all the nodes in the network concurrently exchange packets
of information (such are all the systems in \cite{watts}, for
example), while $1/L$ measures the efficiency of a {\em sequential
system\/} (i.e. only one packet of information goes along the
network). $1/L$ is a reasonable approximation of $\EGLOB$ when
there are not huge differences among the distances in the graph,
and this can explain why $L$ works reasonably well in the
unweighted examples of \cite{watts}. But, in general $1/L$ can
significantly depart from $\EGLOB$. For instance, in the Internet,
having few computers with an extremely slow connection does not
mean that the whole Internet diminishes by far its efficiency: in
practice, the presence of such very slow computers goes unnoticed,
because the other thousands of computers are exchanging packets
among them in a very efficient way. Here $1/L$ would give a number
very close to zero (strictly 0 in the particular case when a
computer is disconnected from the others and $L=+\infty$), while
$\EGLOB$ gives the correct efficiency measure of the Internet. We
turn now our attention to the local properties of a network. $C$
is only one among the many possible intuitive measures
\cite{newman3} of how well connected a cluster is. 
It can be shown that when in a graph most of its local subgraphs $\bf
G_i$ are not sparse, then $C$ is a good approximation of $\ELOC$.
Summing up there are not two different kinds of analysis to be 
done for the global and local scales, but
just one with a very precise physical meaning: the efficiency in
transporting information.

\noindent
We now illustrate the onset of the small-world in an unweighted graph
by means of the same example used in \cite{watts}.
A regular lattice with $N=1000$ and $k=20$
is rewired with probability $p$ and
$\EGLOB$ and $\ELOC$ are reported in fig.1 as functions of $p$
\cite{alg}.
For $p=0$ we expect the system to be inefficient on a global scale
($\EGLOB \sim k/N~log(N/K)$) but locally efficient.
The situation is inverted for the random graph. In fact
at $p=1$ $\EGLOB$ assumes a maximum value of $0.4$, meaning $40\%$
the efficiency of the ideal graph with an edge between each couple
of vertices. This at the expenses of the fault tolerance
($\ELOC \sim 0$).

\begin{figure}
\begin{center}
\epsfig{figure=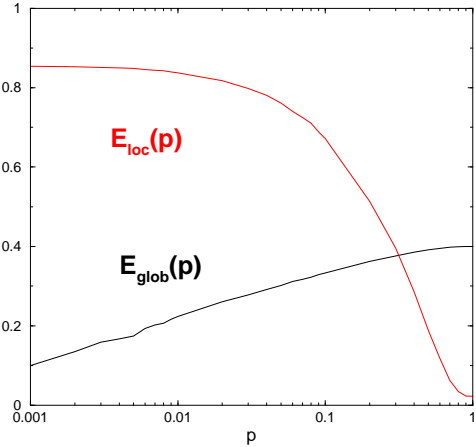,width=0.7\columnwidth,angle=270}
\end{center}
\caption{
FIG.1 Global and local efficiency for the graph example
considered in \protect\cite{watts}. A regular lattice
with $N=1000$ and $k=20$ is rewired with probability $p$.
The small-world behaviour results from the increase of $\EGLOB$
caused by the introduction of only a few rewired edges (short cuts),
which on the other side do not affect $\ELOC$.
At $p \sim 0.1$, $\EGLOB$ has almost reached
the value of the random graph,
though $\ELOC$ has only diminished by very little
from the value of $0.82$ of the regular lattice.
Small worlds have high $\EGLOB$ and $\ELOC$.
}
\end{figure}

The small-world behaviour appears for intermediate values of $p$.
It results from the fast increase of $\EGLOB$
(for small p we find a linear increase of $\EGLOB$ in the logarithmic horizontal scale)
caused by the introduction of only a few rewired edges (short cuts),
which on the other side do not affect $\ELOC$.
At $p \sim 0.1$, $\EGLOB$ has almost reached
the maximum value of $0.4$, though $\ELOC$ has only diminished by
very little from the maximum value of $0.82$.
For an unweighted case the description in terms of network efficiency
resembles the approximation given in \cite{watts}. In particular
we have checked that a good agreement with curves
$L(p)$ and $C(p)$ \cite{watts} can be obtained by
reporting $1/\EGLOB(p)$ and $\ELOC(p)$.
Of course in such an example the short cuts connect at almost
no cost vertices that would otherwise be much farther apart
(because $\ell_{ij}=1 ~\forall i \neq j$).
On the other hand this is not true when we consider a weighted
network.

\noindent
As real networks we consider
first different examples of natural systems (neural networks),
and then we turn our attention to man-made communication and
transportation systems.

{\it 1) Neural Networks.} Thanks to recent experiments
neural structures can be studied at several levels of scale.
Here we focus first on the analysis of the neuroanatomical
structure of cerebral cortex, and then on a simple
nervous system at the level of wiring between neurons.
\noindent
The anatomical connections between cortical areas
are of particular importance for their
intricate relationship with the functional
connectivity of the cerebral cortex \cite{sporns}.
We analyze two databases of cortico-cortical connections
in the macaque and in the cat \cite{scannell1}.
Tab.1 indicates the two networks are small-worlds~\cite{alg}:
they have high $\EGLOB$, respectively $52\%$ and $69\%$
the efficiency of the ideal graph with an edge between
each couple of vertices (just slightly smaller than the best
possible values of $57\%$ and $70\%$ obtained in random graphs)
and high $\ELOC$, respectively $70\%$ and $83\%$,
i.e.\ high fault tolerance ~\cite{fault}.
These results indicate that in neural cortex each region
is intermingled with the others and has grown following a
perfect balance between local necessities (fault tolerance)
and wide-scope interactions.
Next we consider the neural network of C.\ elegans,
the only case of a nervous system completely mapped
at the level of neurons and chemical synapses \cite{white}.
Tab.1 shows that this is also  a small-world network:
C.\ elegans achieves both a $50\%$ of global and local efficiency.
Moreover the value of $\EGLOB$ is similar to $\ELOC$.
This is a difference from cortex databases
where fault tolerance is slighty privileged with respect
to global communication.

{\it 2) Communication Networks.} We have considered two of the
most important large-scale communication networks present
nowadays: the World Wide Web and the Internet. Tab.2 shows that
they have relatively high values of $\EGLOB$ (slightly smaller than  
the best possible values obtained for random graphs) 
and $\ELOC$. Despite the WWW is a virtual network and the Internet is a physical
network, at a global scale they transport information essentially
in the same way (as their $\EGLOB$'s are almost equal). At a local
scale, the bigger $\ELOC$ in the WWW case can be explained both by
the tendency in the WWW to create Web communities (where pages
talking about the same subject tend to link to each other), and by
the fact that many pages within the same site are often quickly
connected to each other by some root or menu page.

{\it 3) Transport Networks.} Differently from previous databases
the Boston subway transportation system ({\em MBTA\/}) can be
better described by a weighted graph, the matrix $\{\ell_{ij}\}$
being given by the geographical distances between stations. 
If we consider the MBTA as an unweighted graph 
we obtain that it is apparently neither locally 
nor globally efficient (see Tab.3).  
On the other hand, when we take into account the geographical 
distances, we have $\EGLOB=0.63$: this shows the {\em MBTA\/} 
is a very efficient transportation system on a global scale, 
only $37\%$ less efficient than the ideal subway with a 
direct tunnel from each station to the others. 
Even in the weighted case $\ELOC$ stays low ($0.03$), 
indicating a poor local behaviour: 
differently from a neural network the {\em MBTA\/} is
not fault tolerant and a damage in a station will dramatically
affect the connection between the previous and the next station.
The difference with respect to neural networks comes from
different needs and priorities in the construction and evolution
mechanism: when we build a subway system, the priority is given to
the achievement of global efficiency, and not to fault tolerance.
In fact a temporary problem in a station can be solved by other
means: for example, walking, or taking a bus from the previous to
the next station. That is to say, the MBTA is not a {\em closed
system\/}: it can be considered, after all, as a subgraph of a
wider transportation network. This property is of fundamental
importance when we analyze a system: while global efficiency is
without doubt the major characteristic, it is {\em closure\/} that
somehow leads a system to have high local efficiency (without
alternatives, there should be high fault-tolerance). The MBTA is
not a closed system, and thus this explains why, unlike in the
case of the brain, fault tolerance is not a critical issue.
Indeed, if we increase the precision of the analysis and change
the MBTA subway network by taking into account, for example, the
Boston Bus System, this extended transportation system comes back
to be a small-world network ($\EGLOB=0.72$, $\ELOC=0.46$).
Qualitatively similar results, confirming the similarity of
construction principles, have been obtained for other undergrounds
and for a wider transportation system consisting of all the main
airplane and highway connections throughout the
world\cite{marchiori2}. Considering all the transportation
alternatives available at that scale makes again the system closed
(there are no other reasonable routing alternatives), and so
fault-tolerance comes back as a leading construction principle.

Summing up, the introduction of the efficiency measure allows to
give a definition of small-world with a clear physical meaning,
and provides important hints on why the original formulas of
\cite{watts} work reasonably well in some cases, and where they
fail. The efficiency measure allows a precise quantitative
analysis of the information flow, and works both in the unweighted
abstraction, and in the more realistic assumption of weighted
networks. Finally, analysis of real data indicates that various
existing (neural, communication and transport) networks exhibit
the small-world behaviour (even, in some cases, when their
unweighted abstractions do not), substantiating the idea that the
diffusion of small-world networks can be interpreted as the need
to create networks that are both globally and locally efficient.
%
%

\pagestyle{empty}
\begin{table}
\caption{~~~ Macaque and cat cortico-cortical
connections ~\protect\cite{scannell1}.
The macaque database contains $N=69$ cortical areas and $K=413$
connections \protect\cite{young}.
The cat database has $N=55$ cortical areas (including hippocampus,
amygdala, entorhinal cortex and subiculum) and $K=564$
(revised database and cortical parcellation from \protect\cite{scannell2}).
The nervous system of C.\ elegans
consists of $N=282$ neurons and $K=2462$ links
which can be either synaptic connections or
gap junctions \protect\cite{verme}.
\label{table1}}

\begin{tabular}{l|ll|ll|l}
 & $\EGLOB$ & $\ELOC$ \\
\hline
Macaque                  & 0.52 & 0.70\\
Cat                      & 0.69 & 0.83\\
\hline C.\ elegans               & 0.46 & 0.47
\end{tabular}
\end{table}

\pagestyle{empty}
\begin{table}
\caption{~~~ Communication networks. Data on the World Wide Web
 from   http://www.nd.edu/\symbol{126}networks 
contains $N=325729$ documents and $K=1090108$ links
\protect\cite{barabasi1}, while the Internet database is taken
from http://moat.nlanr.net and has $N=6474$ nodes and $K=12572$
links. \label{table2}}

\begin{tabular}{l|ll|ll|l}
 & $\EGLOB$ & $\ELOC$ \\
\hline WWW                    & 0.28 & 0.36\\
Internet                      & 0.29 & 0.26
\end{tabular}
\end{table}

\pagestyle{empty}
\begin{table}
\caption{~~~ The Boston underground transportation system  ({\em
MBTA\/}) consists of $N=124$ stations and $K=124$ tunnels. 
The matrix $\{\ell_{ij}\}$ of the spatial
distances between stations, used for the weighted case, has been
calculated using databases from http://www.mbta.com/ and the U.S.\
National Mapping Division. \label{table2}}

\begin{tabular}{l|ll|ll|l}
& $\EGLOB$ & $\ELOC$ \\
\hline MBTA (unweighted) & 0.10 & 0.006\\
MBTA (weighted) & 0.63  & 0.03
\end{tabular}
\end{table}
\noindent


\begin{thebibliography}{99}
\bibitem{alg}
Here and in the following the matrix $\{d_{ij}\}_{i,j\in G}$
has been computed by using two different methods:
the Floyd-Warshall ($O(N^3)$) \cite{pallottino}
and the Dijkstra algorithm ($O(N^2~log N)$) \protect\cite{newman3}.
\end{thebibliography}
\end{document}